\documentclass[conference]{IEEEtran}
\usepackage{amsmath,amssymb,epsfig,xcolor} % easier math formulae, align, subequation
\usepackage{tabularx,multirow,array}
\usepackage{times}
\usepackage{booktabs}
\usepackage{multirow}
\usepackage{soul}
\usepackage{enumerate}
\usepackage[font=small]{caption}
\usepackage[countmax]{subfloat}
\usepackage{algpseudocode}
\usepackage{algorithm}
\usepackage{subfigure}
\usepackage{cite}
\usepackage{url}
\usepackage{gensymb}
\begin{document}

\title{Machine Learning-Assisted Anomaly Detection in Maritime Navigation Using AIS Data}

\author{
    \IEEEauthorblockN{Sandeep Kumar~Singh and Frank~Heymann}
    \IEEEauthorblockA{German Aerospace Center (DLR), Germany 
    \\\{sandeep.singh, frank.heymann\}@dlr.de}
}

\maketitle
%%\vspace{-1.5cm}
\begin{abstract}
The automatic identification system (AIS) reports vessels' static and dynamic information, which are essential for maritime traffic situation awareness. However, AIS transponders can be switched off to hide suspicious activities, such as illegal fishing, or piracy. Therefore, this paper uses real world AIS data to analyze the possibility of successful detection of various anomalies in the maritime domain.   We propose a multi-class artificial neural network (ANN)-based anomaly detection framework to classify intentional and non-intentional AIS on-off switching anomalies. The multi-class anomaly framework  captures AIS message dropouts due to various reasons, e.g., channel effects  or intentional  one for carrying illegal activities. We extract position, speed, course and timing information from real world AIS data, and use them to train a 2-class (normal and anomaly) and a 3-class (normal, power outage and anomaly) anomaly detection models. Our results show that the models achieve around 99.9\% overall accuracy, and are able to classify a test sample in the order of microseconds.
\end{abstract}
%Notably, our model can be applied to any other advance mo
%%\vspace{-0.3cm}
%\begin{IEEEkeywords}
%%\vspace{-0.3cm}
%Elastic optical networks, spectrum allocation, fragmentation, blocking analysis, approximation.
%\end{IEEEkeywords}

%\vspace{-0.3cm}
\section{Introduction}
\par Maritime security is of utmost importance today as over  90\% international trade as well as thousands of passengers around the world are carried over sea \cite{riveiro2018maritime}.   Maritime transport poses significant challenges, natural as well as human-induced, e.g., tough and unpredictable environment,  collision,  illegal fishing, smuggling, pollution, and piracy. Nowadays, the automatic identification system (AIS) has become an essential part of maritime traffic situation awareness. Vessels equipped with AIS transponders report their positions, which are based on the global navigation satellite system (GNSS), their navigational status, as well as other voyage related information. This information can be used in collision-avoidance mechanism, tracking of vessels, detecting unusual trajectories of vessels, etc. However, AIS transponders can be switched off to hide suspicious activities, such as illegal fishing, or piracy. Therefore, it is essential to use real world AIS data in order to analyze the possibility of successful detection of various anomalies in the maritime domain. 

Anomaly detection in maritime traffic has attracted researchers to apply various statistical and machine learning solutions \cite{tu2017exploiting,riveiro2018maritime}. Statistical solutions, such as extended Kalman filter and particle filter, have been used  to reconstruct trajectory of vessels \cite{perera2012maritime}. When the estimated and real trajectories differ more than a predefined threshold, those events are categorized as an anomalous behavior. Bayesian network was applied  for tackling missing values (anomalies) in a dataset for prediction and classification purposes \cite{Hruschka2007}. A hidden Markov model was utilized to detect AIS on-off  switching (OOS) anomaly with the consideration of transmission channel characteristics \cite{guerriero2010analysis}.
Machine learning (ML) algorithms, such as artificial neural network (ANN), support vector machine (SVM), long short-term memory (LSTM), on the other hand, have been shown to perform better than statistical methods for prediction and classification problems in general. Recently, \cite{mazzarella2017novel} proposed a one-class SVM-based anomaly detection framework that takes AIS data as well as received signal strength into consideration for analyzing AIS OOS anomaly. \cite{zhong2019vessel} used a random forest ML algorithm to classify vessels using AIS data streams. A long-short-term-memory (LSTM) algorithm was used in \cite{yuan2019novel} to reconstruct vessels' trajectories. Although the use of ML models for anomaly detection has just started recently, ANN, which can learn complex fitting functions and has been shown promising as compared to other ML techniques,  has not been applied to deal with the multi-class intentional and non-intentional AIS OOS anomaly yet.  A multi-class model is vital to capture AIS message dropouts due to various reasons, for example,  channel effects (e.g., power outage) or intentional (for illegal activities).  We expect more AIS dropouts, when vessels move away from the AIS receiver. Thus, it is important to distinguish between the intentional and natural AIS OOS. The natural dropouts of AIS messages can be identified using the vessel’s distance from an AIS receiver where the weak received signal strength may result in loss of messages. 

In this paper, we propose an ANN-based anomaly detection framework to detect an AIS OOS anomaly.  We use real-world AIS messages to train our multi-class  anomaly detection framework, and test on further real AIS data.  More in detail, we  extract a four-dimensional (4-D) feature vector containing latitude, longitude, speed, and course  information from each AIS message transmitted by vessels. After resampling the received AIS messages, we train the ANN models with data samples  containing one or more 4-D feature vectors within an observation period. A data sample is labeled as an anomaly when it has consecutive AIS dropouts more than a predefined threshold. First we train and test a 2-class model consisting of normal and anomaly samples.  Our results show that the 2-class AIS OOS model can detect anomaly correctly with an overall accuracy of nearly 100\%. In order to distinguish intentional AIS OOS anomaly from the AIS dropouts due to weak received signal strength,  we add “power outage” as another class.  For this purpose we add the distance as an extra feature, calculated between the position of vessels and a receiver station, to classify power outage samples based on a predefined AIS transmission range. Upon training the 3-class anomaly model with this additional feature, the model achieves around 99.9\% overall accuracy, and that too in a reasonable time of  a few microseconds per sample.  

\par The rest of this paper is organized as follows.  Section \ref{sec:framework} briefly describes the AIS and its usefulness in the anomaly detection framework. Section \ref{sec:ANN} presents a neural network model for anomaly detection and data preparation steps for the models. We evaluate the performance of the models in Section \ref{sec:evaluation}, and conclude the paper in Section \ref{sec:conclusion}.

%We resample the received AIS messages from each vessel at every $\Delta$ seconds to fit them according to their timestamps. Sixty of these consecutive 4-D feature vectors are then arranged sequentially to form a data sample to train the ANN model. As the number of anomalies in the real AIS data is generally small,  we artificially create new anomalies by dropping some of the AIS messages from each vessel's trajectory. When a data sample has more than X\% (X is a design parameter) of consecutive AIS dropouts, then we label that sample as an anomaly, otherwise normal. Our results show that the 2-class AIS OOS model can detect anomaly correctly with true positive rate of 97\%, which means the false negative rate (i.e., classifying anomaly sample as normal) is very low (3\%). Moreover, the false positive rate (classifying normal as anomaly) is even lower (1\%). As the number of normal data samples is higher than the anomaly samples, the model obtains overall accuracy as 99\%.

%%\vspace{-0.3cm}
\begin{figure*}[ht!]
 \centering
\includegraphics[width=0.8\textwidth]{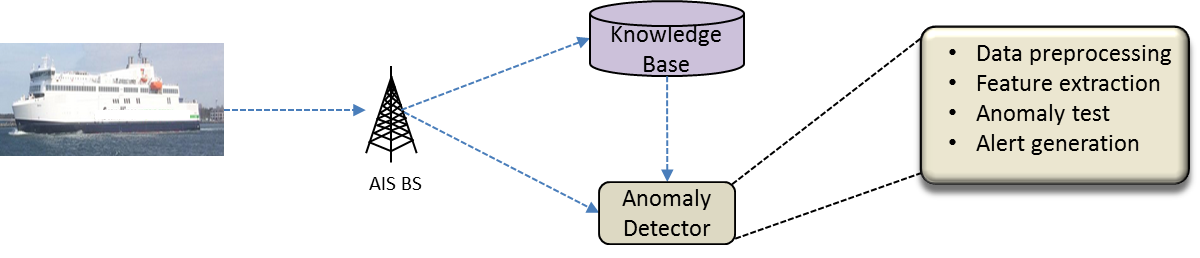}
%\vspace{-0.2 cm}
  \caption{An anomaly detection framework.}
% \vspace{-0.4cm}
\label{fig:framework}
\end{figure*}
\section{AIS and Anomaly Detection}\label{sec:framework}
In this section, we first briefly describe the AIS, and then we explain our ML-aided framework that can be used to detect various maritime anomalies, including the AIS on-off switching, using the AIS data. 

\subsection{AIS for Maritime Security}
Today most ships, excluding very small non-passenger ones (less than 300 gross tons), are mandated to be equipped  with AIS transponders by the international maritime organization, and follow the Safety Of Life At Sea (SOLAS) convention, 2002, in order to ensure security and safety of maritime navigation \cite{solas2003international}. 
AIS transponders mainly use two dedicated wavelengths in a very-high frequency (VHF) band to transmit AIS data: 161.975MHz and 162.025MHz. Among different classes of transponders, Class A is of particular interest in this paper as most of the ships are legally required to install it. The AIS transmitted data are received by GNSS antennas. Both terrestrial as well as satellite links have been used for the AIS transmission. Note that a terrestrial link is limited by the curvature of the Earth, roughly 40 nautical miles, in the normal propagation condition. 

AIS transponders can transmit 27 types of messages, but most used messages are of type position reports (1, 2, 3, and 5). An AIS position report message transmitted by a vessel at regular interval contains mainly vessel identification number (MMSI), longitude (lon), latitude (lat), speed over ground (SOG), course over ground (COG), heading, rotation rate, timestamp, and other voyage related information. The position report messages, which are our  focus,  are sent every 2 to 10 seconds depending on a vessel's speed, and every 3 minutes for anchored or moored vessels. The Class A AIS transponders use Self Organized Time Division Multiple Access (SOTDMA) protocol to transmit AIS information in self- managed time slots. 
The positional and navigational information make AIS data a treasure for detecting anomaly behavior of vessels, predicting trajectory, avoiding collisions, among others. In this paper, we store position report messages from vessels with speed larger than 3 knots, and remove data corresponding to anchored or moored vessels, by checking  their speed  and navigational status. We use them to train our ML model in the anomaly detection framework, which is described next. 
\subsection{ML-Assisted Anomaly Detection Framework}
Figure \ref{fig:framework} depicts our anomaly detection framework, which uses AIS messages from vessels as input data.  Upon reception of an AIS message through the base station (BS), a Knowledge Base is used to store past AIS information in order to train ML-assisted anomaly detector. The trained ML model is saved and updated regularly in the Knowledge Base for detecting anomalies on live AIS data traffic. The anomaly detector has the following roles to play.
\begin{itemize}
\item \textbf{Data preprocessing}. The AIS transmission is generally affected by propagation phenomena, including transmission frequencies, power, antenna gains, and most  importantly intentional switching on and off of AIS on-board transponders \cite{iphar2015detection}. Therefore,  at the receiver AIS messages containing dynamic report about vessels' position, speed, course, and other voyage details  could be missing or erroneous when time sampled at an expected  interval (e.g., every 2 seconds for moving vessels with speed greater than 23 knots). The missing AIS messages are also called dropouts in the literature \cite{mazzarella2017novel}. For a sequential  data analysis, we can use either dummy or interpolated data to replace missing messages. The erroneous/corrupted AIS messages must be dealt carefully and is out of scope of this paper.

\item \textbf{Feature extraction}. Depending on the type of anomaly, a set of features can be extracted from each AIS message for the training and testing of an anomaly detector. In this paper, for AIS on-off switching, we consider mainly position, COG and SOG features of a vessel. 

\item \textbf{Anomaly test}. The anomaly detector can use a suitable  supervised or unsupervised ML algorithm for training and testing purposes. In our case we use a supervised ANN model to detect whether the received data over a time is anomalous or not. The ANN model is trained with data collected over time in the Knowledge Base, and used for online anomaly detection purposes.   

\item \textbf{Decision making}. As various factors impact the reception of AIS messages, a decision making block can further post process the result obtain from an anomaly test. It can decide and differentiate whether the detected anomaly is  natural, e.g., power outage, or an intentional one, for example, illegal fishing or piracy. 

\end{itemize}  

\section{ML Framework and Data Preparation}\label{sec:ANN}
\subsection{Artificial Neural Network }
\begin{figure}[ht!]
 \centering
\includegraphics[scale=0.8]{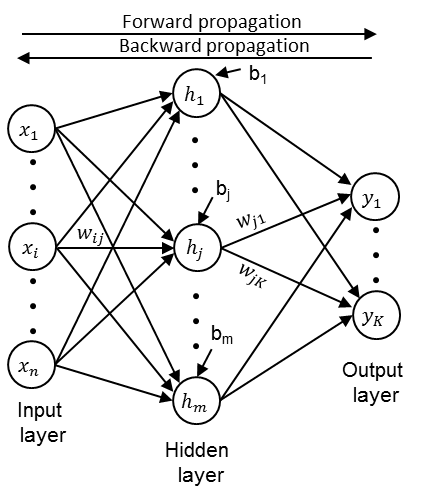}
%\vspace{-0.2 cm}
  \caption{ Artificial neural network classifier model.}
% \vspace{-0.4cm}
\label{fig:ANN}
\end{figure}

We use a three-layer ANN model, shown in Figure \ref{fig:ANN}, which has an input layer with $n$  nodes,
a hidden layer with $m$ nodes, and an output layer with $K$ nodes. The weight of the link between an input node $i$ and a hidden node $j$ is denoted as $w_{ij}, i = 1, \ldots, n, j = 1, \ldots, m$, and the link-weight between the hidden node $j$ and an output node corresponding to a class $k, k=1,2, \ldots, K$ is denoted as $w_{jk}$. The output of a hidden node $j$ is expressed as
\begin{equation}
h_j = f\left(\sum_{i=1}^n w_{ij}x_i - b_j \right), j = 1, 2, \ldots, m.
\end{equation} 
where $b_j$ is the threshold of the hidden layer node $j$, and $f(\cdot)$ is an activation function that transforms its input on a non-linear output function. In the hidden layer,  we use a Rectified Linear Unit (ReLU) as an activation function due to its efficient computation and better gradient propagation. The output of the ReLU function given an input $x$ is  $\max(0, x)$. In the output layer, the activation function is Softmax, which is expressed as 
\begin{equation*}
\hat{y}_k = f(z_k) = \frac{e^{z_k}}{\sum_{j=1}^K e^{z_j}}\cdot
\end{equation*}
The Softmax function calculates the probabilities of each target class $k=1, 2, \ldots, K$ over all possible target classes. Finally, error is calculated between the actual output per class, $y_k$ and the estimated output $\hat{y}_k$  using the cross entropy error or loss function as given below. 
\begin{equation*}
e = - \sum_{k=1}^K y_k \log(\hat{y}_k)
\end{equation*}
Each training epoch transfers the signal information from input layer to the output layer, and computes the error. A regularization term is generally added to the loss function to avoid overfitting of training data. Although there is no analytical formula which can provide the optimum weights and biases to get minimum error,  an iterative backpropagation algorithm \cite{hecht1992theory} propagates the error backwards  to adjust the weights of links and biases. Among the many optimizers available today for ANN, we selected a well-known Adam optimizer  which finds the (local) minimum of  the objective (error) function making use of its gradient. The process is repeated until the error is below a defined threshold or the number of epochs is reached to its maximum defined limit. In general, we can design the ANN with one or more hidden layers and neurons. It is important to mention here that for our experiment increasing the number of hidden layers increases the learning time complexity, but accuracy.

\subsection{Data Preparation}
\begin{figure}[ht!]
 \centering
\includegraphics[width=0.45\textwidth]{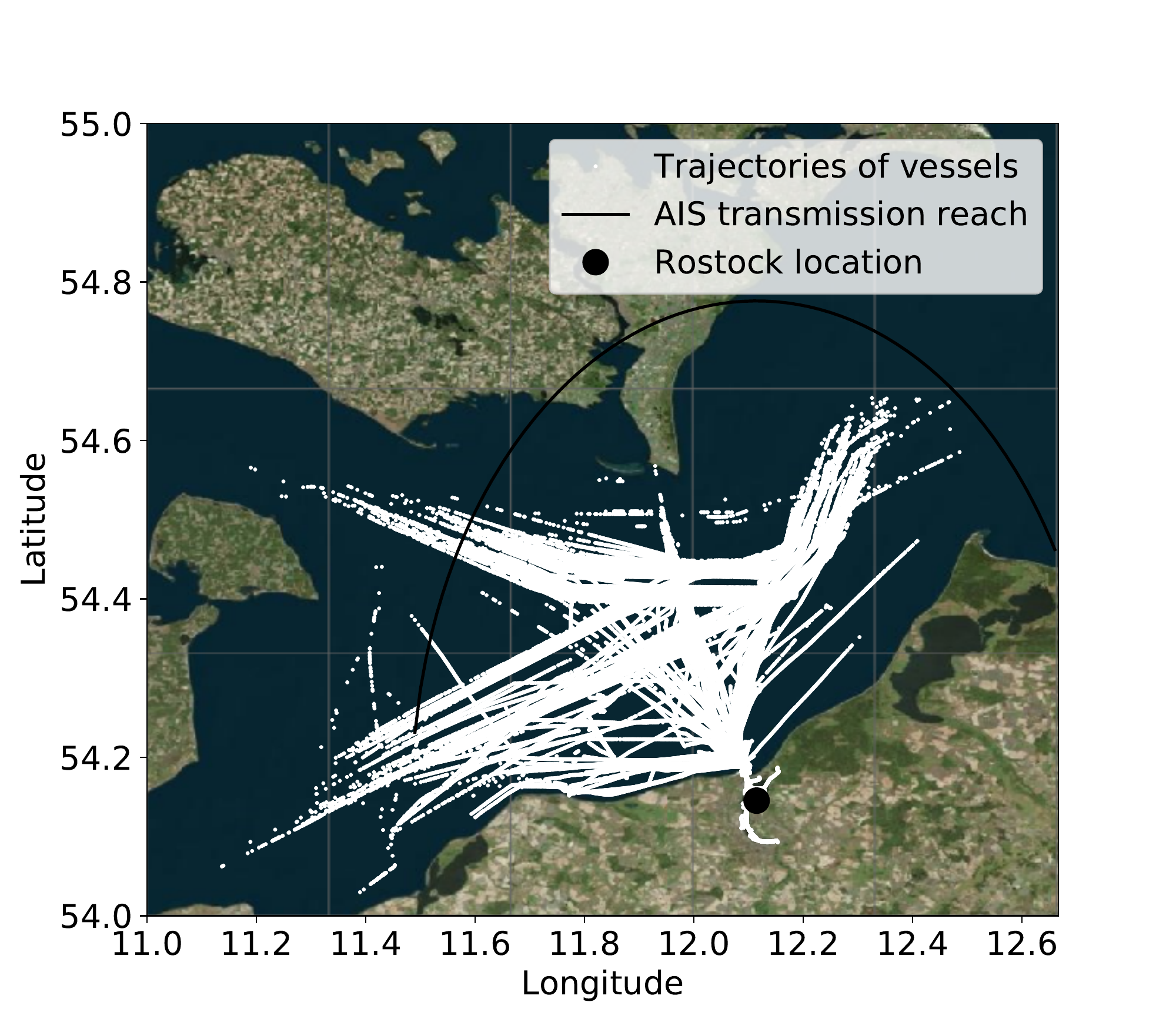}
%\vspace{-0.2 cm}
  \caption{ Vessels' trajectories extracted from AIS data.}
% \vspace{-0.4cm}
\label{fig:all_trajecories}
\end{figure}

We use the AIS data collected at the Rostock port. After decoding the AIS messages, we extract trajectories of vessels from AIS data using their unique MMSIs, and associated longitude and latitude information. Figure \ref{fig:all_trajecories} depicts trajectories of vessels that have been extracted from the received AIS data. It also shows the considered boundary for transmission reach under the normal condition , i.e., distance around 40 nautical miles, from Rostock. We can see that there are many missing data samples. Moreover, the AIS transmission rate vary between 2 and 10 seconds for moving  vessels with speed more than 3 knots.  Therefore, for each trajectory we resample the data every $\tau = 2$ seconds  and  place the real data according to their timestamps, and missing data by a dummy value.  We denote a sequence of time-sampled data within an observation period $T=120$ seconds as an input data sample to the ANN, which is represented by $X_{t_i} =\{x_{t_i}, x_{t_i +\tau}, \ldots, x_{t_i + (T-1)\tau}\}$, where the first data $x_{t_i}$ is a real 4-dimensional vector derived from a reported AIS message of a vessel at time $t_i$. 

\begin{equation*}
x_{t_i} \equiv \{lat, lon, COG, SOG\} 
\end{equation*}

On the other hand, $x_{t_i +\tau}$ could be a real 4-D feature vector  received at time $t_i +\tau$ or  dummy values representing a missing data.
For example, with $\tau =2$ seconds, and $T = 120$, we have an input data sample of size $T/\tau = 60$.  We use a sliding window to create  data samples from each vessel as $\{X_{t_i}, X_{t_{i+1}}, \ldots \}$, where $t_i $ is the time instance at which $i^{th}$  AIS message was received. To begin with the AIS on-off detection, each data sequence is labeled by one of two classes: normal or anomaly. Later we add another subclass of anomaly, which is non-intentional, i.e., AIS dropouts due to power outages. For the 2-class anomaly detection model, we  label a data sample as an anomaly if it has more than $X\%$ continuous AIS dropouts following the first AIS data in a sample. As the number of anomaly samples is usually much lower than normal samples, which will  have negative effect on the learning process, we synthetically create more anomaly samples by appending  236 (59 times 4) missing values to the 4-D feature of each real AIS messages of  vessels in order to balance the number of samples per class in the dataset.   Similarly, for the 3-class model we first synthetically generated AIS anomalies as in the 2-class model, and some of which could be power outage samples. We then created extra power outage samples by modifying the positions of vessels with uniformly distributed random positions in the regions away from Rostock, which is located at (longitude = 12.1\degree, latitude = 54.1\degree). Precisely,  we selected two rectangular areas with coordinates (long, lat in degrees) as [(11, 54), (11.4, 55)] and [(12.4, 54.4), (13, 55)], such that distances of these artificial positions are more than the AIS transmission range, i.e., 40 nautical miles. It is important to mention that the AIS transmission range depends on the propagation condition, atmospheric effect, antenna heights, transmission power, etc. Nevertheless, it is generally considered as 40 nautical miles under normal propagation condition, which we use it as a threshold to define power outage samples among the anomaly samples.  Table \ref{tab:data_samples} summarizes the number of real and synthetic  data samples for the normal, anomaly and power outage classes in both the models.
 
% \usepackage{multirow}
\begin{table}[]
\centering
\caption{The number of real and synthetic samples per class}
\label{tab:data_samples}
\begin{tabular}{|l|c|c|c|c|c|}
\hline
\multirow{2}{*}{} & Normal & \multicolumn{2}{c|}{AIS-OOS anomaly} & \multicolumn{2}{c|}{Power outage} \\ \cline{2-6} 
                  & Real   & Real       & Synthetic       & Real          & Synthetic         \\ \hline
2-class           & 129,470 & 1,875       &   132,352         &               &                   \\ \hline
3-class           & 129,470 &  1,725          &   141,048              &     150          &   123,152                \\ \hline
\end{tabular}
\end{table}

In total, we collected 165,704 AIS messages  from 228 stationary and moving vessels in the Baltic sea for three days from the Rostock AIS base station.  From these messages, we use 132,352 messages of 133 vessels with speed more than 3 knots to train and validate the ANN models, and data of remaining vessels were kept aside for testing the model. Note that not all real AIS messages forms real samples in the dataset. The reason is that some messages (1007=132,352-129,470-1875) overlaps with other real messages in the same 2 seconds resampling interval. In the 2-class model, the dataset eventually has 129,470  normal  samples and 134,227 anomaly samples (which includes synthetically created anomalies).  In the 3-class model, the number of normal, anomaly and power outage samples are  129,470, 142,773, and 123,302, respectively. For both of these models, we randomly split the datasets into training and validation purposes with the ratio 60:40. The testing is also performed with data collected from vessels that are not used in training.

\section{Performance Evaluation}\label{sec:evaluation}
\par In this section, we investigate the effectiveness of our neural network-assisted anomaly detection framework in terms of accuracy, loss, computation time, and a confusion matrix. A confusion matrix is a performance measure of a classification algorithm that shows  the number of correct and incorrect prediction of samples for each class. In a 2-class (normal and anomaly) model, true positive (TP) rate is the probability of correct detection of anomaly, and true negative (TN) rate is the right recognition of the normal behavior. Therefore, false positive (FP) rate is 1-TN, and false negative (FN) rate is 1-TP. On the other hand, in a model with 3 or more classes, one-vs-rest approach can be used to compute correct and false prediction. The overall accuracy is defined as the ratio of total number of correct predictions and total number of input samples. We train and validate the ANN models with 10 different seeds, which randomly splits the dataset into 60:40 in each run. Thus, the accuracy presented on all (ten times) validation samples. We use a Scikit-learn library's implementation of ANN and the default parameters provided by it.  The ANN is fully connected with one hidden layer with 100 neurons. We observe that increasing hidden layers does not improve accuracy. We use a well-known Adam optimizer. The learning rate is adaptive and starting learning rate is 0.001. The convergence is considered to be reached when the loss does not improve by at least the tolerance $10^{-5}$ for a certain number of consecutive iterations. The threshold for defining the anomaly in terms of fraction of missing data in a sample is set to $X=$ 0.9, i.e., 90\%, which means an anomaly sample has 54 or more AIS 4-D feature vector missing after the first AIS 4-D feature vector. Unless otherwise stated, an output neuron with the highest probability is defined as a predicted class. 

%In other words,  TP  and FN can be defined as how many samples for a class $k$ is predicted correctly, and falsely as other classes, and how many samples of other classes is predicted as the class $k$.

\begin{figure}[ht!]
 \centering
\includegraphics[clip, trim=5cm 10cm 5cm 10cm, width=0.45\textwidth]{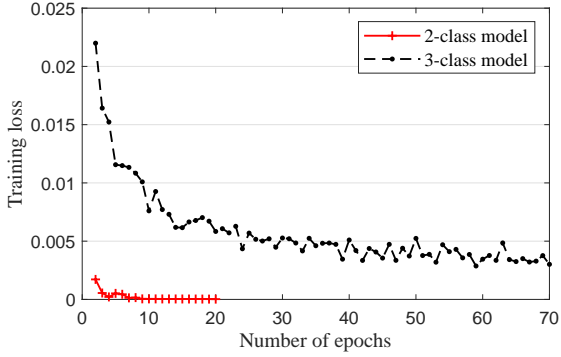}
%\vspace{-0.2 cm}
  \caption{Losses during training of the ANN 2-class and 3-class models are shown against the number of training epochs.}
% \vspace{-0.4cm}
\label{fig:loss}
\end{figure}
Let us first investigate how well the ANN models have been trained. Figure \ref{fig:loss} illustrates the cross-entropy loss (i.e, error in estimating the classes) of both 2-class and 3-class models during the training process after each epoch (iteration). Note that the lower loss refers to a better learned model, and we can see that within 20 epochs the 2-class model reaches the convergence, and  the final loss is in the order of $10^{-5}$. This means the 2-class model optimizes the model weights to its near-optimal values and could estimate with higher accuracy than the 3-class model. The loss in 3-class model takes more number of epochs (70) to  converge, and the final loss is in the order of  $10^{-3}$. It is also important to note that we run the experiments on a commodity computer with a 2-core CPU, 8 GB RAM and Intel i7 processor, and the training times per 1000 samples of the 2-class and 3-class models are 0.24 and 3.18 seconds, respectively. The testing times per 1000 samples in the 2-class and 3-class models are 0.004 and 0.12 second, respectively. Thus, we can say that the ANN anomaly models can be used in real time to monitor live maritime traffic to detect anomalies. 

%Figure \ref{fig:execution-time} illustrates the execution times of the ANN model (log-scale) during training and testing with varying number of neurons per hidden layer.  We observe that the training time of ANN is longer than the testing time for the same number of data samples. More importantly, with a 10-fold (linear) increase in the number of neurons per hidden layer, the execution time increases quadratically. Additionally, the training and test execution times increase with the increase in the number of hidden layers. For all cases of number of neurons and hidden layers shown here, the performance of ANN in terms of accuracy of anomaly detection is above 99\%, which is remarkable. 

\begin{figure}[ht!]
 \centering
\includegraphics[clip, trim=0cm 16cm 18cm 0cm, width=0.4\textwidth]{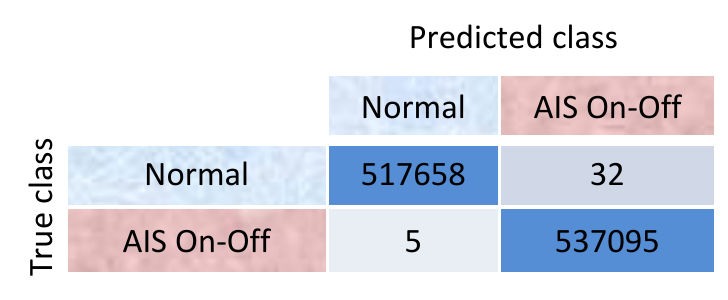}
%\vspace{-0.2 cm}
  \caption{A confusion matrix of normal and anomaly classes.}
% \vspace{-0.4cm}
\label{fig:CM1}
\end{figure}

\begin{figure}[ht!]
 \centering
\includegraphics[clip, trim=0cm 15cm 15cm 0cm, width=0.49\textwidth]{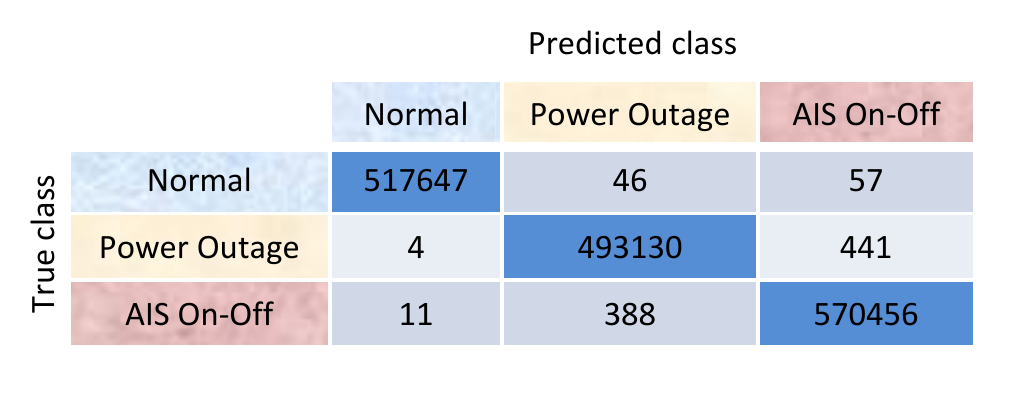}
%\vspace{-0.2 cm}
  \caption{A confusion matrix of normal, power outage, and AIS on-off anomaly classes.}
% \vspace{-0.4cm}
\label{fig:CM2}
\end{figure}

Next, we evaluate the performance of both the models in terms of accuracy, which is computed  with 10 different seeds. Each run has different training (60\%) and validation (40\%) data samples, and the accuracy is obtained on the validation data samples. Figure \ref{fig:CM1} depicts the performance of the 2-class ANN model in terms of a confusion matrix. We can see that the model can correctly learn normal as well as anomaly behavior of the vessels with an overall accuracy (number of true predictions divided by number of samples) of nearly 100\%.  Moreover, only 32 normal samples are falsely predicted as anomaly samples, so FP is negligible, and FN (i.e., predicting anomaly as normal) is also negligible.  In order to distinguish the power outage and AIS on-off anomaly, Figure \ref{fig:CM2} shows the the confusion matrix obtained using the 3-class ANN model. Interestingly, the 3-class model  misclassifies more data samples than the 2-class model. In particular, the power outage samples misclassify its  4 and 441 samples as normal and AIS-OOS anomaly, respectively. Similarly,  11 and 338 AIS-OOS anomaly samples are incorrectly classified as normal and power outage samples, respectively.   We observe that the misclassification rate between power outage and AIS-OOS anomaly classes is around 0.1\%, which occur mainly at the  transmission range boundary (i.e., 40 nautical miles).  Nevertheless, it also achieves  99.9\% accuracy. This shows that our ML-assisted anomaly framework is able to learn the distinction between the AIS OOS due to power outage and the intentional ones. 

\begin{figure}[ht!]
 \centering
\includegraphics[width=0.49\textwidth]{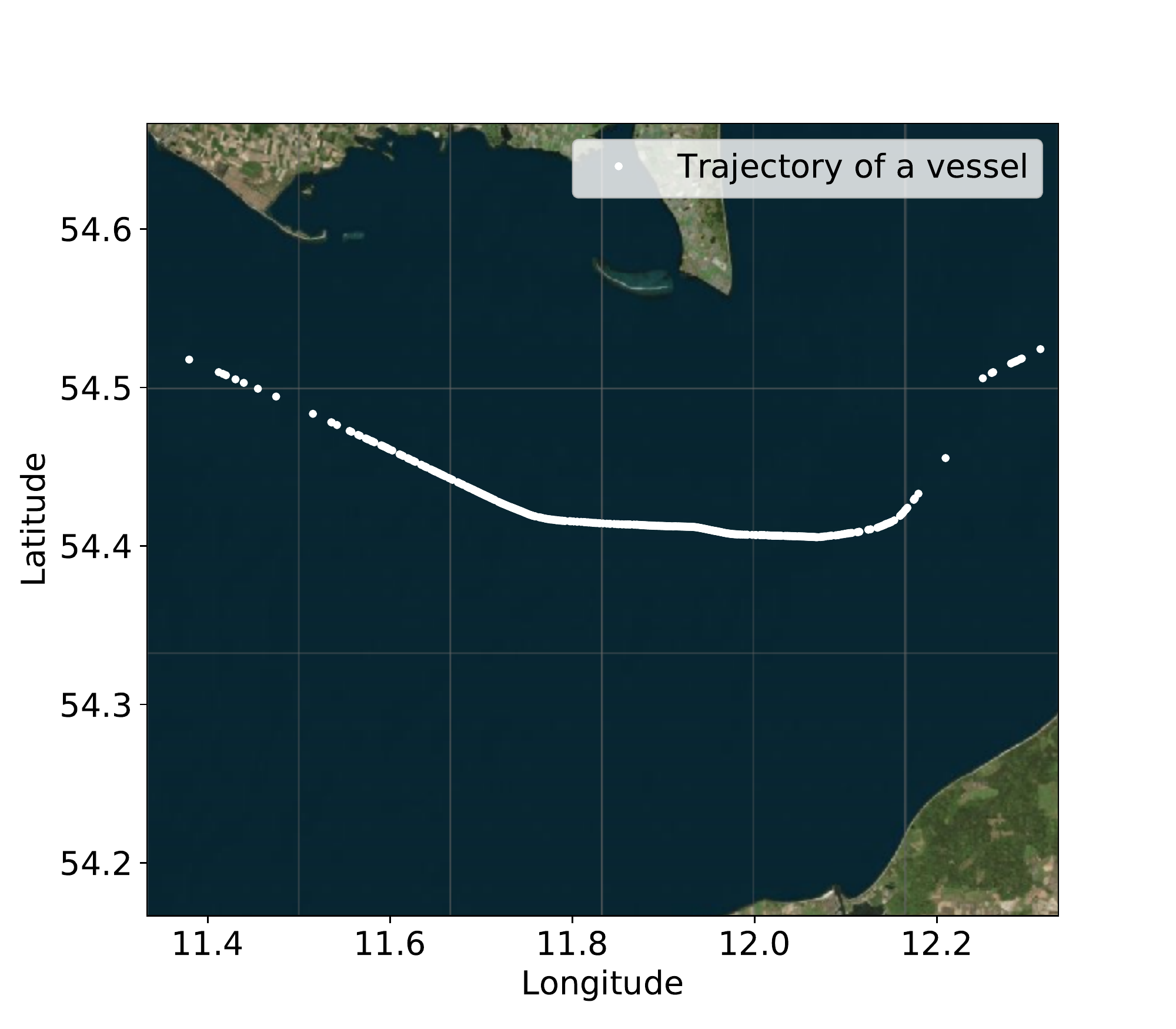}
%\vspace{-0.2 cm}
  \caption{A vessel's trajectory.}
% \vspace{-0.4cm}
\label{fig:track}
\end{figure}

\begin{figure}[ht!]
 \centering
\includegraphics[width=0.49\textwidth]{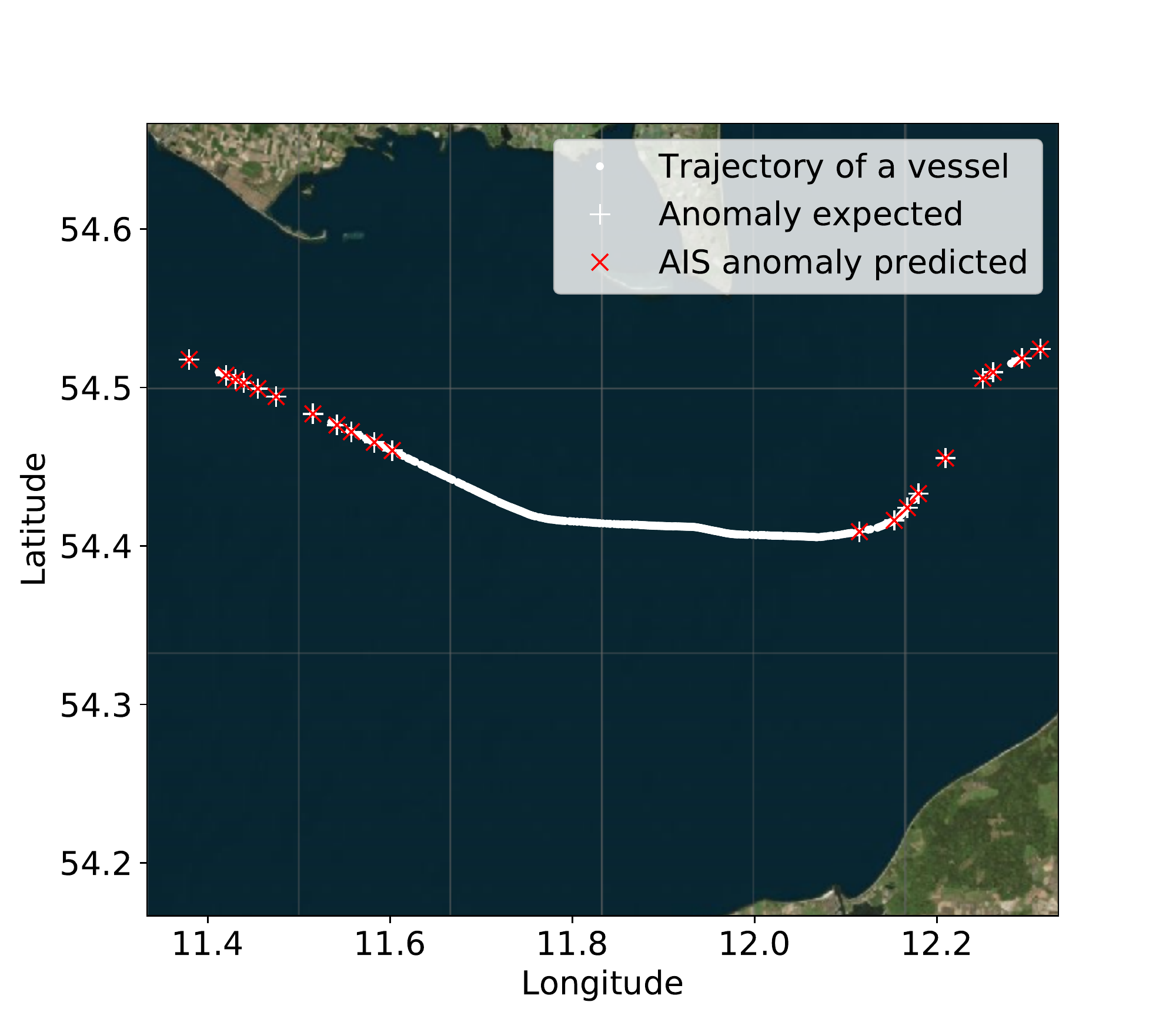}
%\vspace{-0.2 cm}
  \caption{Anomaly detection in a vessel's trajectory.}
% \vspace{-0.4cm}
\label{fig:anomaly-track}
\end{figure}

\begin{figure}[ht!]
 \centering
\includegraphics[width=0.49\textwidth]{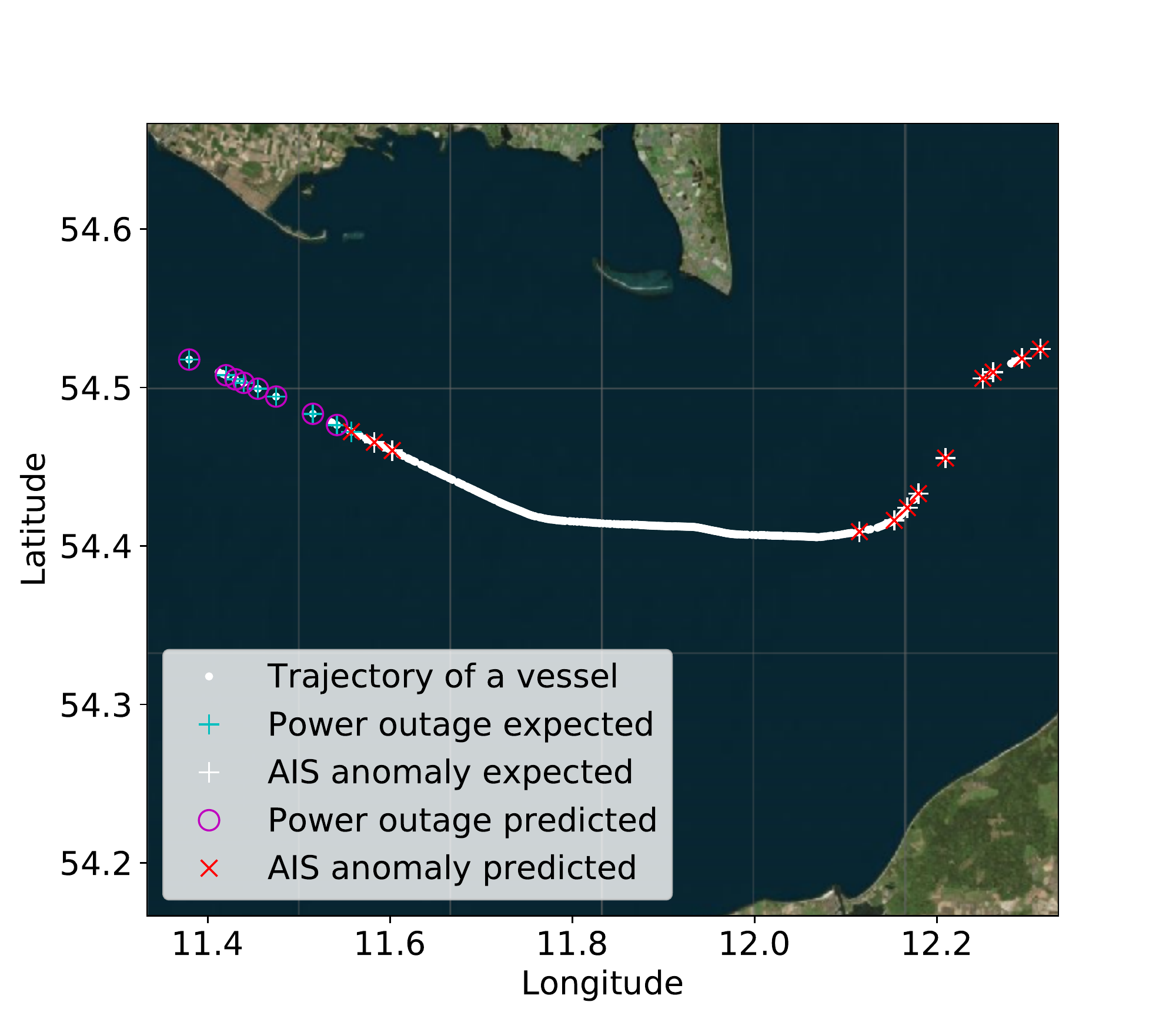}
%\vspace{-0.2 cm}
  \caption{Anomaly and power outage detection in a vessel's trajectory.}
% \vspace{-0.4cm}
\label{fig:anomaly-po-track1}
\end{figure}

Finally, we evaluate the performance of both models on a single vessel's AIS data, which is not used in either training or validation of the models. The trajectory formed by the vessel's AIS position (latitude and longitude) data is shown in Figure \ref{fig:track}. We can see the discontinuities in the trajectory on both edges of the track. In fact, there are many missing AIS data  on both edges. In order to see the anomalies, in Figure \ref{fig:anomaly-track} we highlight coordinates where we expect anomalies as defined in the trained ANN models. More precisely, the anomaly coordinates are marked with plus sign where there is no consecutive AIS data for more than 90\% (108 seconds) of observation period (120 seconds) after receiving the first AIS data in each sample.  The figure also illustrates the predicted anomalies by the 2-class model. Figure \ref{fig:anomaly-track} shows that the 2-class anomaly model can correctly identify all anomalies. 

\begin{figure}[ht!]
 \centering
\includegraphics[width=0.49\textwidth]{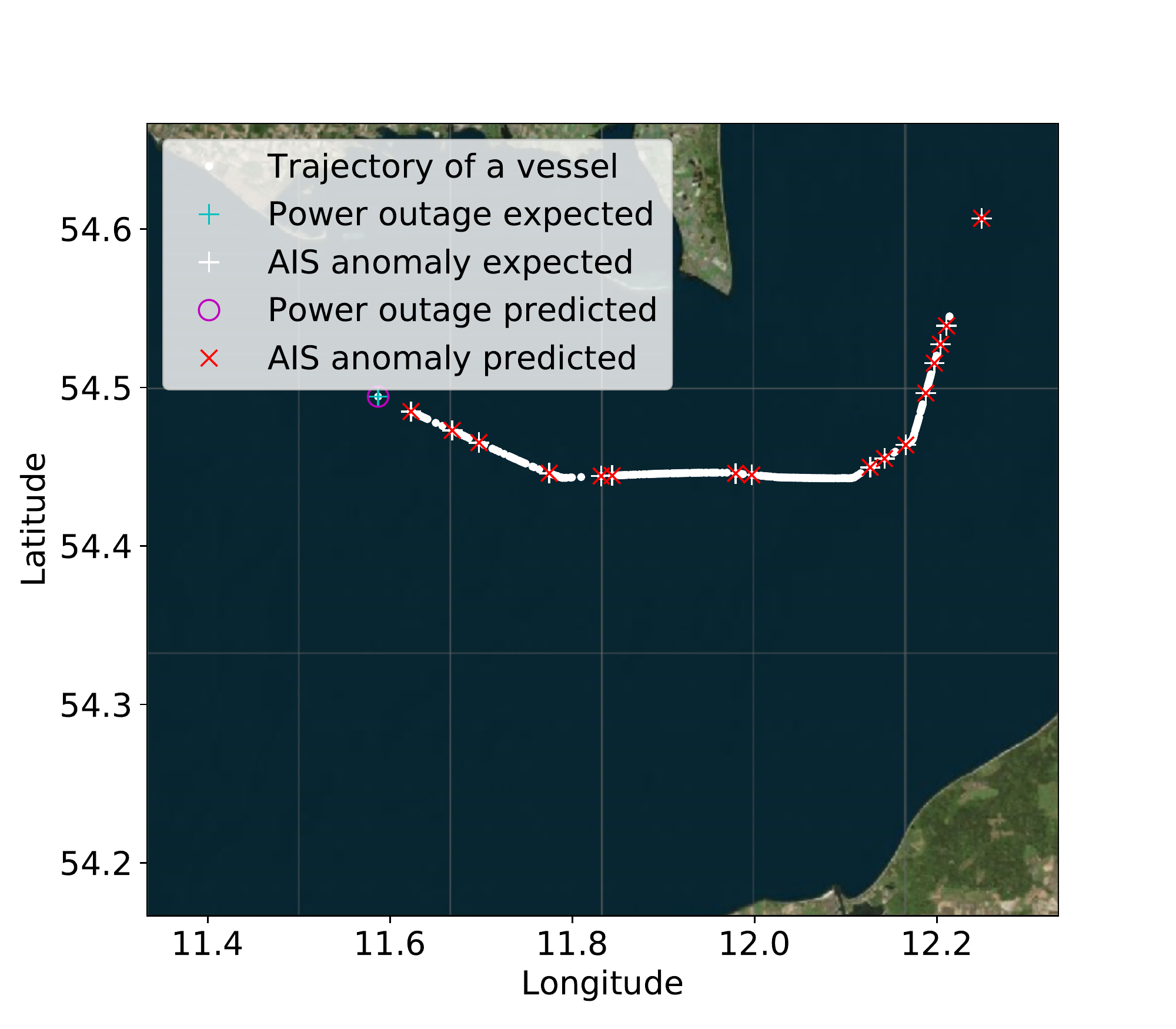}
%\vspace{-0.2 cm}
  \caption{Anomaly and power outage detection in a vessel's trajectory.}
% \vspace{-0.4cm}
\label{fig:anomaly-po-track2}
\end{figure}

\begin{figure}[ht!]
 \centering
\includegraphics[clip, trim=0cm 15cm 15cm 0cm, width=0.49\textwidth]{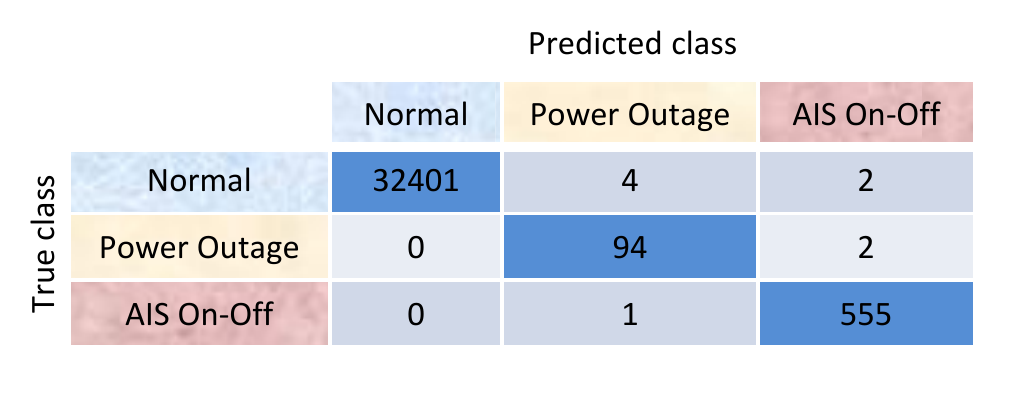}
%\vspace{-0.2 cm}
  \caption{A confusion matrix of normal, power outage, and AIS on-off anomaly classes based on a new real dataset.}
% \vspace{-0.4cm}
\label{fig:CM3}
\end{figure}

As mentioned before, some of these anomalies might occur due to power outages as the vessel move beyond the AIS transmission reach. Therefore, we used the 3-class model to predict normal, power outage and anomaly samples in Figure \ref{fig:anomaly-po-track1}. We observe here that the model correctly classifies all three classes, except that one expected power outage sample is misclassified as an anomaly sample. The reason is that this sample is around the boundary of AIS transmission reach. However, this is not the case with another track, shown in Figure \ref{fig:anomaly-po-track2}. In contrast to the previous figure, here we observe an instance where a power outage sample at the boundary is correctly predicted.  In order to see the correct and false predictions in all the tracks whose AIS data were not used  in either training or validation, Figure \ref{fig:CM3} shows the confusion matrix on new real AIS data from 71 vessels. As only a negligible amount of misclassification occurs, we can say that the 3-class model is able to learn the normal, power outage and OOS anomaly characteristics from the AIS data.

%\begin{figure}[ht!]
% \centering
%\includegraphics[width=0.49\textwidth]{Anomaly_all_new.pdf}
%%\vspace{-0.2 cm}
%  \caption{Anomaly and power outage detection in new vessels' trajectories.}
%% \vspace{-0.4cm}
%\label{fig:anomaly-all}
%\end{figure}

\section{Conclusions} \label{sec:conclusion}  
\par In this paper,  we proposed  a multi-class artificial neural network (ANN)-based anomaly detection framework to classify intentional and non-intentional AIS on-off switching anomaly. The multi-class anomaly model  captures AIS message dropouts due to channel effects  and intentional  one. We utilized position, speed and course, and timing information from real world AIS data for the training the multi-class anomaly detection models. Our results show that the models achieve around 99.9\% overall accuracy, and misclassify only a few samples.
The anomaly detection framework is designed in such a way that we can further classify the anomaly types, for instance, anomaly when vessels are moored, make U-turns, or enter into a restricted zone. For the future work, we would apply our models to live maritime traffic scenarios. 
%%\vspace{-0.3cm}
\bibliographystyle{IEEEtran}
\bibliography{AnomalyBib}

\end{document}